\documentclass[11pt]{article}
\usepackage[numbers, sort&compress, square]{natbib}  

\usepackage[labelfont=bf]{caption} 
\usepackage[font={small}]{caption}   

\usepackage{amsmath}
\usepackage{amssymb}
\usepackage{url}
\usepackage{placeins}
\usepackage{textcomp}
\usepackage{graphicx}
\usepackage{float} 
\usepackage{authblk}
\usepackage{chngpage}
\usepackage{lineno}

\begin{document}
\title{\textbf{Demonstration of a quantized microwave quadrupole insulator with topologically protected corner states}}

\author[1]{Christopher W. Peterson}
\author[2]{Wladimir A. Benalcazar}
\author[2]{\mbox{Taylor L. Hughes}}
\author[3,$\ast$]{Gaurav Bahl}
\affil[$$]{\footnotesize{University of Illinois at Urbana-Champaign}}
\affil[1]{\footnotesize{Department of Electrical and Computer Engineering}}
\affil[2]{\footnotesize{Department of Physics}}
\affil[3]{\footnotesize{Department of Mechanical Science and Engineering}}
\affil[$\ast$]{\footnotesize{To whom correspondence should be addressed; bahl@illinois.edu}}

\date{\today}

\maketitle

\textbf{\small{
	The modern theory of electric polarization in crystals associates the dipole moment of an insulator with a Berry phase of its electronic ground state \cite{zak1989,vanderbilt1993}.
	This concept constituted a breakthrough that not only solved the long-standing puzzle of how to calculate dipole moments in crystals, but also lies at the core 
	of the theory of topological band structures in insulators and superconductors, including the quantum anomalous Hall insulator \cite{thouless1982,chang2013} and the quantum spin Hall insulator \cite{kane2005,bernevig2006,konig2007}, as well as quantized adiabatic pumping processes \cite{king-smith1993,thouless1983,fu2006}.
	A recent theoretical proposal extended the Berry phase framework to account for higher electric multipole moments \cite{benalcazar2017quad}, revealing the existence of topological phases that have not previously been observed. 
	Here we demonstrate the first member of this predicted class -- a quantized quadrupole topological insulator -- experimentally produced using a GHz-frequency reconfigurable microwave circuit.
	We confirm the non-trivial topological phase through both spectroscopic measurements, as well as with the identification of corner states that are manifested as a result of the bulk topology.
	We additionally test a critical prediction that these corner states are protected by the topology of the bulk, and not due to surface artifacts, by deforming the edge between the topological and trivial regimes.
	Our results provide conclusive evidence of a unique form of robustness which has never previously been observed. 
} }

\vspace{12pt}

\begin{figure}
	\begin{adjustwidth}{-1in}{-1in}
		    \centering
		    \makebox[\textwidth][c]{\includegraphics[scale = 0.6]{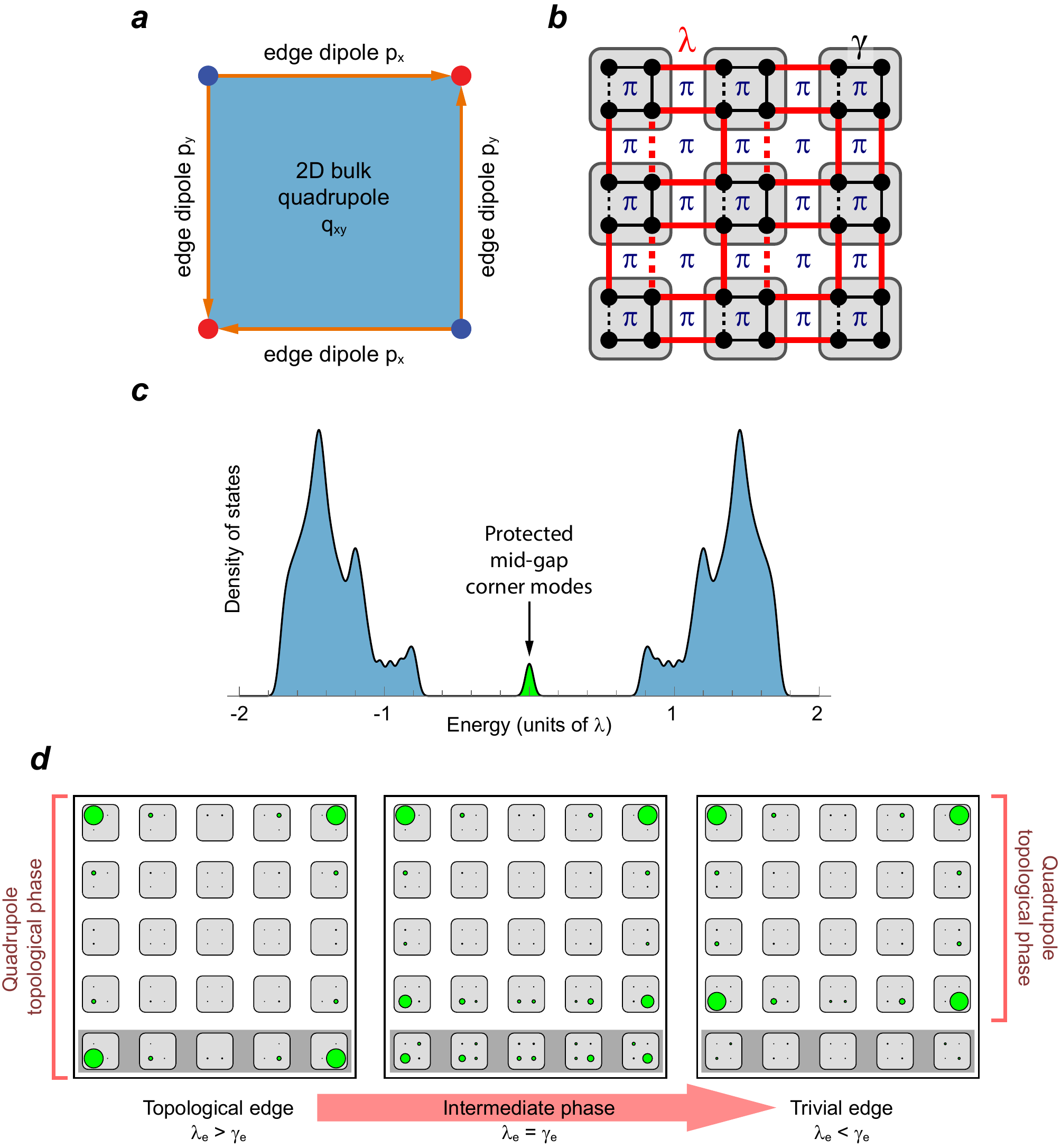}}%
		    \caption{
\textbf{ Quadrupole topological  insulator -- } 
\textbf{(a)} A 2D bulk quadrupole topological insulator (blue square) manifests edge-localized topological dipoles (orange lines) and corner-localized charges of $\pm e/2$ (red and blue dots, respectively).
\textbf{(b)} The tight-binding representation of the model that realizes a quadrupole topological insulator having four sites per unit cell. $\lambda$ are couplings between unit cells and $\gamma$ are couplings within unit cells. Dashed lines indicate a $-1$ phase factor on the coupling, a gauge choice for the creation of a synthetic magnetic flux of $\pi$ per plaquette. The model is in the quadrupole topological phase for $\lambda>\gamma$ and in the trivial phase for $\lambda<\gamma$.
\textbf{(c)} Theoretically calculated density of states for the quantized quadrupole insulator ($5 \times 5$ unit cells) shown in (b) with full open boundaries. The lower and upper bands (blue) have eigenstates delocalized in the bulk, while the states in the middle of the gap (green) are localized on the corners, as shown in (d).
\textbf{(d)} Theoretically calculated probability density functions of the four in gap modes as the unit cells on the lowest edge are reconfigured from $\lambda_e / \gamma_e = 4.5$ (left) to $\lambda_e / \gamma_e = 1$ (center), and to $\lambda_e / \gamma_e= 1/4.5$ (right). Throughout the deformation only $\gamma_e$ is changed, $\lambda_e = \lambda = 1$ and $\gamma=1/4.5$ in all plots. This deformation is experimentally tested in Fig.~\ref{fig4}.
}
	\label{fig1}
	\end{adjustwidth}
\end{figure}

The simplest model of a system with a quantized dipole moment is a two-band insulator in 1D \cite{ssh1979} which, due to the presence of chiral or inversion symmetries~\cite{hughes2011inversion,turner2011}, exhibits quantized fractional edge charges of $\pm e/2$ when its band structure is topological.
Microscopically, the fractional edge charges of the quantized dipole insulator are associated with a pair of edge-localized bound states of the Hamiltonian.
These edge states have energies that lie within the bulk insulating gap and have been observed in 1D lattices in systems of cold atoms \cite{leder2016,meier2016} as well as in several metamaterial contexts \cite{krauss2012,slobozhanyuk2015,blanco-redondo2016,chaunsali2017}.
However, the possible existence of quantized higher multipole moments in crystalline insulators has remained an outstanding problem for the past 25 years.
A theory addressing this problem was recently put forth in Ref. \cite{benalcazar2017quad}, in which a pair of simple electronic lattice models were proposed in 2D and 3D that exhibit the signatures of quantized electric quadrupole and octupole moments, respectively.
A 2D insulator with a quantized quadrupole moment $q_{xy}=e/2$ generates \emph{edge-localized} dipole moments tangent to the edge, and \emph{corner-localized} charges, both of magnitude $e/2$ (Fig.~\ref{fig1}a).
Microscopically, the corner charges are associated with four corner-localized modes that lie in the middle of the energy gap \cite{benalcazar2017quad, benalcazar2017quadPRB} (Fig.~\ref{fig1}c).
The edge-localized polarizations arise from the gapped, but topological nature of the edge states themselves, however, they do not have a spectroscopic manifestation.

Metamaterial analogues of the quantum Hall and quantum spin Hall topological insulators have previously been implemented in phononic and photonic systems \cite{Hafezi_top12, Rechtsman13, Irvine, Huber_helix15}.
These implementations are not quantum mechanical, but preserve coherence and hence can exhibit topological properties.
In this paper, we implement the precise 2D quadrupole topological model from Ref. \cite{benalcazar2017quad} (Fig.~\ref{fig1}b) in a metamaterial composed of coupled microwave resonators.
Although the edge polarizations and the corner-localized energy modes are both signatures that owe their existence to the non-trivial topology of the bulk energy bands, we focus on the latter, as they can provide direct spectroscopic evidence for the existence of the non-trivial quadrupole topological phase.
Specifically, we experimentally demonstrate the existence of mid-gap energy modes that are localized at the corners of the lattice (Fig.~\ref{fig1}d, left).
Furthermore, we provide evidence that these corner modes are not due to surface effects, but are required by the topological bulk phase.
We accomplish this by deforming one of the edges from the topological to trivial regime, and we observe that the mid-gap corner modes are not destroyed; instead, they recede inward into the sample to the corners on the newly generated boundaries of the quadrupole topological phase (Fig.~\ref{fig1}d). 

\vspace{12pt}

\begin{figure}
	\begin{adjustwidth}{-1in}{-1in}
		    \centering
		    \makebox[\textwidth][c]{\includegraphics[scale = 0.6]{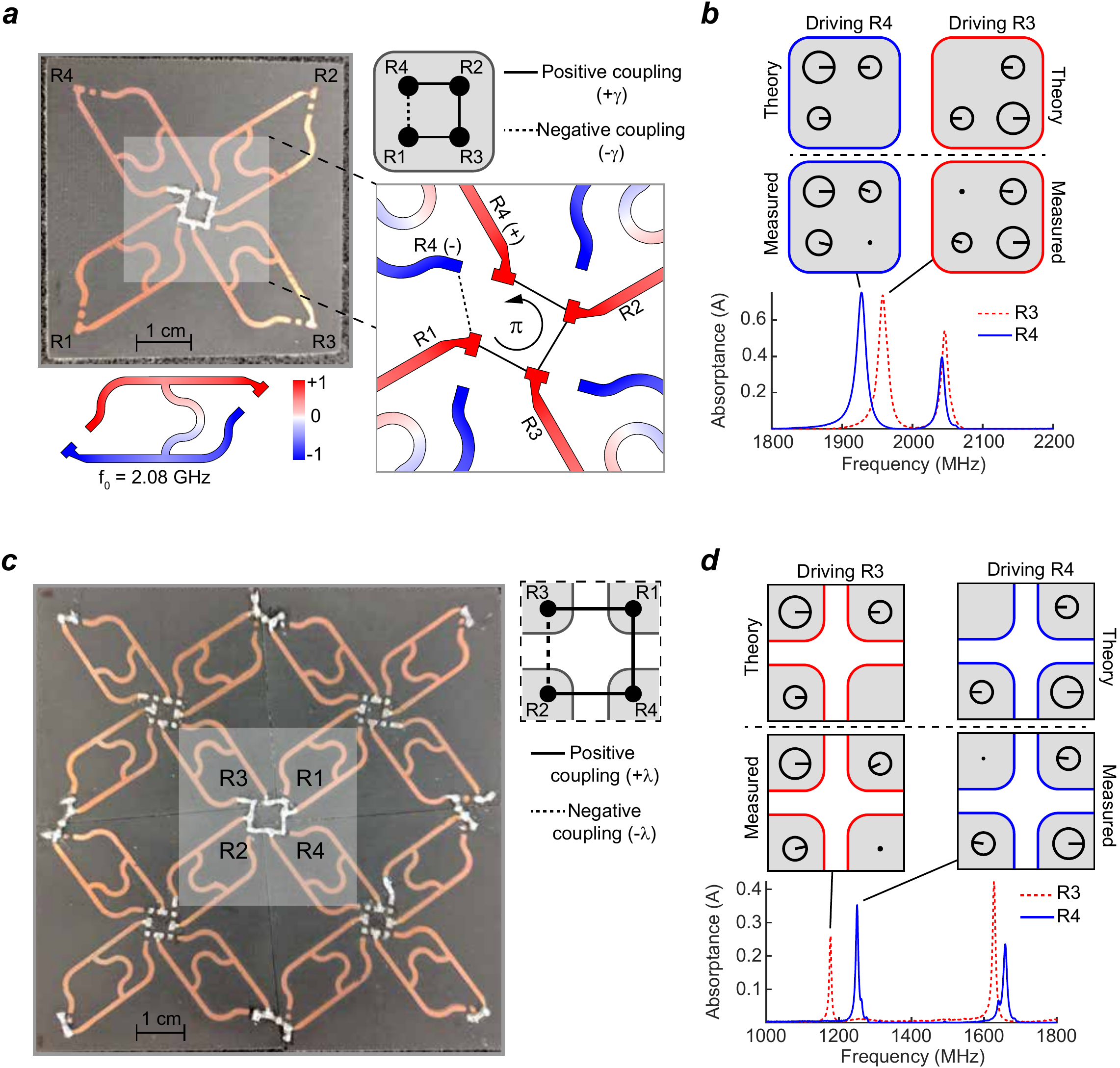}}%
		    \caption{
		    %
		    \textbf{Verification of microwave quadrupole lattice bulk topology -- }
		    \textbf{(a)} A unit cell of the quadrupole topological insulator (photograph on top-left) is composed of four capacitively coupled H-shaped microstrip resonators, each having a fundamental mode at 2.08~GHz as illustrated (colors represent voltage amplitude). The coupling between R4 and R1 is set as negative, as shown in the detailed schematic, in order to produce $\pi$ flux through the unit cell plaquette. $\gamma \approx 35$~MHz.
		    %
		    \textbf{(b)} Eigenmode verification for the unit cell plaquette -- The resonator frequencies are shifted to $\sim2$~GHz due to capacitive loading from the couplers. The theoretical and measured eigenmodes are presented as phasor diagrams; circle diameter corresponds to the magnitude of the resonator excitation, the line corresponds to the phase (0 is to right, increases counter-clockwise). When driving R4, resonator R1 is seen to be in-phase, confirming the negative coupling between resonators R1 and R4.
		    \textbf{(c)} A 2x2 test array of unit cells with $\gamma \rightarrow 0$. Negative coupling is set between resonators R2 and R3 as illustrated in the schematic. $\lambda \approx 150$~MHz.
		    \textbf{(d)} Eigenmodes for the plaquette formed by the 4 central resonators of a 2x2 array are similar as to those of the unit cell due to the $\pi$ flux. When driving R3, resonator R2 is seen to be in-phase, confirming the negative coupling between these resonators. Here, the resonator frequencies are shifted to $\sim1.4$~GHz due to greater capacitive loading.
		    }
	\label{fig2}
	\end{adjustwidth}
\end{figure}

The microwave quadrupole topological insulator we implement consists of a square lattice of unit cells, where each unit cell is composed of four identical resonators (Fig.~\ref{fig1}b). 
The coupling rates $\gamma$ and $\lambda$ describe coupling between resonators within the same unit cell and between adjacent unit cells, respectively.
Each plaquette, a square of any four adjacent resonators, contains a single negative coupling term (dashed lines in Fig.~\ref{fig1}b) which amounts to the generation of a synthetic magnetic flux of $\pi$ puncturing the plaquette. 
The existence of this non-zero flux opens both the bulk and edge energy gaps, which are necessary to protect the corner-localized mid gap modes.
Each resonator in our experimental array is implemented using an $H$-shaped microstrip transmission line that has a fundamental resonance at $f_0 = 2.08$~GHz, having typical linewidth $\sim 20$~MHz, with a spatial voltage distribution as illustrated in Fig.~\ref{fig2}a (bottom).  
At the center of the cross piece lies a voltage node while each end-point of the $H$-shape is a quarter-wavelength from the center and is therefore an anti-node.
Adjacent tips of the $H$ are separated by a half-wavelength and thus differ in phase by $\pi$ rad, and the pinched-in ends are designed to bring the anti-nodal points having opposing phase physically close together.
The unique resonator geometry facilitates the coupling of adjacent resonators either in-phase (positive coupling) or out-of-phase (negative coupling).
To produce the quadrupole topology, in each plaquette we arrange three couplings as positive and one coupling as negative as shown in Fig.~\ref{fig2}a.
We first experimentally confirm that a $\pi$ flux threads each plaquette by examining the limiting cases of $\lambda \rightarrow 0$ and $\gamma \rightarrow 0$.
This experimental verification of $\pi$ flux is necessary to ensure that the spectral features we measure are due to the bulk quadrupole topology, as corner modes themselves, even topologically protected ones, are not unique to the quadrupole topological insulator \cite{benalcazar2017quadPRB,teo2013existence,benalcazar2014classification}.
In the $\lambda \rightarrow 0$ limit, the array consists of isolated unit cell plaquettes as shown in Fig.~\ref{fig2}a.
The expected behavior can be predicted theoretically by a direct diagonalization of the 4-site Hamiltonian whose tight-binding representation is shown by the gray unit cell of Fig.~\ref{fig2}a.
For a coupling rate $\gamma$ between all resonators, and a $\pi$ flux threading the plaquette, the eigenfrequencies are $\pm \sqrt{2}\gamma$, and each of these are two-fold degenerate (see Supplement \S S1).
Since the non-trivial topology of the full array is fully manifest in either the upper or lower band, we choose to characterize only the lower band at $-\sqrt{2}\gamma$.
The measured power absorptance spectrum (the ratio of absorbed power to incident power) of an isolated unit cell is shown in Fig.~\ref{fig2}b -- details on the measurement technique are discussed in Methods. 
As predicted, we find two pairs of nearly degenerate modes separated by $2 \sqrt{2} \gamma \approx 100$~MHz.
The spatial distribution of the lower pair of modes is measured through the voltage amplitude and phase response at each resonator within the plaquette when either resonator R3 or R4 is stimulated (see Methods).
We find good agreement between the magnitudes \emph{and} phases of the theoretical and measured modes (Fig.~\ref{fig2}b).
Characteristic mode shapes appear due to destructive interference, caused by the $\pi$ flux, between counter-circulating paths around the plaquette.
Specifically, when R4 is excited the mode vanishes on the diagonal resonator R3 (and vice versa).
In Supplement \S S1, we discuss the clear contrast of this observation against the anticipated modes of plaquettes having zero flux, although the cases can exhibit spectral similarities.
In the $\gamma \rightarrow 0$ limit, the array consists of isolated inter-unit cell plaquettes (Fig.~\ref{fig2}c, highlighted region).
This plaquette is nearly identical to the isolated unit cell, the differences being that the negative coupling is placed between R2 and R3 and the coupling rate ($\lambda$) is larger.
We experimentally verify that the eigenmodes of this inter-unit cell plaquette also have the features expected for $\pi$ flux by performing similar measurements to the single unit cell case (Fig.~\ref{fig2}d).
For this measurement, the capacitors that originally coupled the resonators within the unit cells are removed to ensure $\gamma = 0$.
We also find good agreement between the theoretical and measured mode shapes, although the lower pair of modes are not perfectly degenerate due to asymmetric capacitive loading (see Supplement \S S2).
The measured frequency separation ($2 \sqrt{2}\lambda \approx 430$~MHz) between the two pairs of modes is approximately $4.3$ times larger than in the isolated unit cell, revealing that ratio of coupling rates is $\lambda / \gamma \approx 4.3$.

\vspace{12pt}

\begin{figure}
	\begin{adjustwidth}{-1in}{-1in}
		    \centering
		    \makebox[\textwidth][c]{\includegraphics[scale = 0.6]{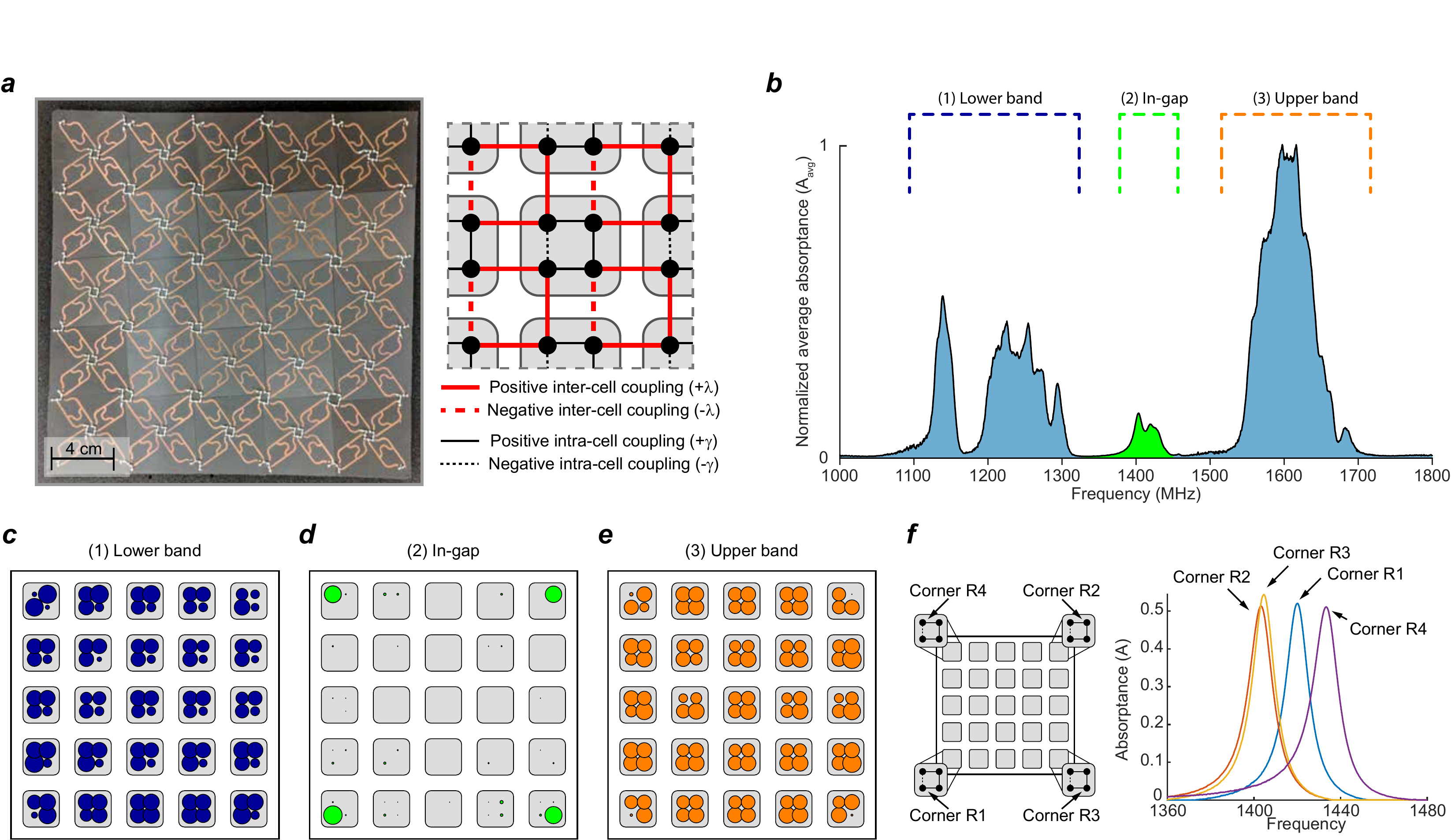}}%
		    \caption{
		    \textbf{Demonstration of microwave quadrupole topological insulator -- }
		    \textbf{(a)} Photograph of the experimental array of coupled resonators that form the quadrupole lattice. The array has $5 \times 5$ unit cells (Fig.~\ref{fig2}a). We set the couplings to a ratio $\lambda / \gamma \approx 4.3$. 
		    The schematic shows the connectivity of a bulk unit cell.
		    \textbf{(b)} Normalized average absorptance spectrum (ratio of absorbed microwave power to incident power) of all the resonators in the array (see Methods for details). We observe two large bands (blue) separated by a band gap containing in-gap modes (green).
		    \textbf{(c)} Spatial distribution of absorptance summed over the lower frequency band indicated in (b). Within this band, the response is dominated by bulk and edge resonators. Circle areas are correspond to local absorptance.
		    \textbf{(d)} Spatial distribution of absorptance summed over the in-gap band indicated in (b). The in-gap modes are localized only on the corner resonators, which are not excited in the lower or upper band.
		    \textbf{(e)} Spatial distribution of absorptance summed over the upper band indicated in (b), which again shows excitation of the bulk and edge resonators.
		    \textbf{(d)} Individual absorptance spectra of the corner resonators reveal that each corner resonator only supports a single mode. 
		    %
		    }
     \label{fig3}
	\end{adjustwidth}
\end{figure}

With the experimental verification of the local plaquette building blocks in place, we construct a quadrupole topological insulator using a $5 \times 5$ array of unit cells (Fig.~\ref{fig3}a) having coupling ratio $\lambda/\gamma \approx 4.3$ and the topology described in Fig.~\ref{fig1}b.
The power absorptance spectrum of each resonator in the full array is measured in the same way as in the isolated plaquettes. The average absorptance across the entire array is presented in Fig.~\ref{fig3}b.
Three spectral bands are identifiable: broad lower and upper bands (blue) separated by a bandgap, and a narrow band of modes near the center of the bandgap (green).
The spatial distributions of each of these bands, obtained by summing over each band indicated in Fig.~\ref{fig3}b, are shown in Figs.~\ref{fig3}c, \ref{fig3}d, and \ref{fig3}e, respectively.
We find that, as predicted in Ref. \cite{benalcazar2017quad}, modes in the lower and upper bands are predominantly localized on the bulk and edge resonators.
The modes in the center of the bandgap, associated with corner charges in the case of an electrical insulator, are highly localized on the corner resonators only. 
In Fig.~\ref{fig3}f we examine the measured spectra within the bulk bandgap of the individual corner resonators, revealing that each corner supports only a single mid-gap mode.
As a consequence of disorder in the array, which breaks chiral and reflection symmetries, the measured spectrum is asymmetric with respect to its mid-gap point. 
Two main sources of disorder exist: (i) systematic differences in the capacitive loading of resonators within the array, and (ii) random disorder due to small manufacturing variations in the capacitance of the discrete coupling capacitors (see discussion in Supplement \S S2.2).
The main spectroscopic effect of the systematic disorder is a splitting of the lower band, which manifests in isolated plaquettes as a lifting of the degeneracy of the lower pair of modes (Fig.~\ref{fig2}).
Despite such disorder and asymmetries, we find that the robust spectral features of the quadrupole topological insulator remain, e.g. the spectral bands are gapped, with only 4 resonances at positions close to mid-gap. Furthermore, we have verified that these four mid-gap modes are tightly confined to the corners (Fig.~\ref{fig3}e).

\vspace{12pt}

\begin{figure}
	\begin{adjustwidth}{-1in}{-1in}
		    \centering
		    \makebox[\textwidth][c]{\includegraphics[scale = 0.6]{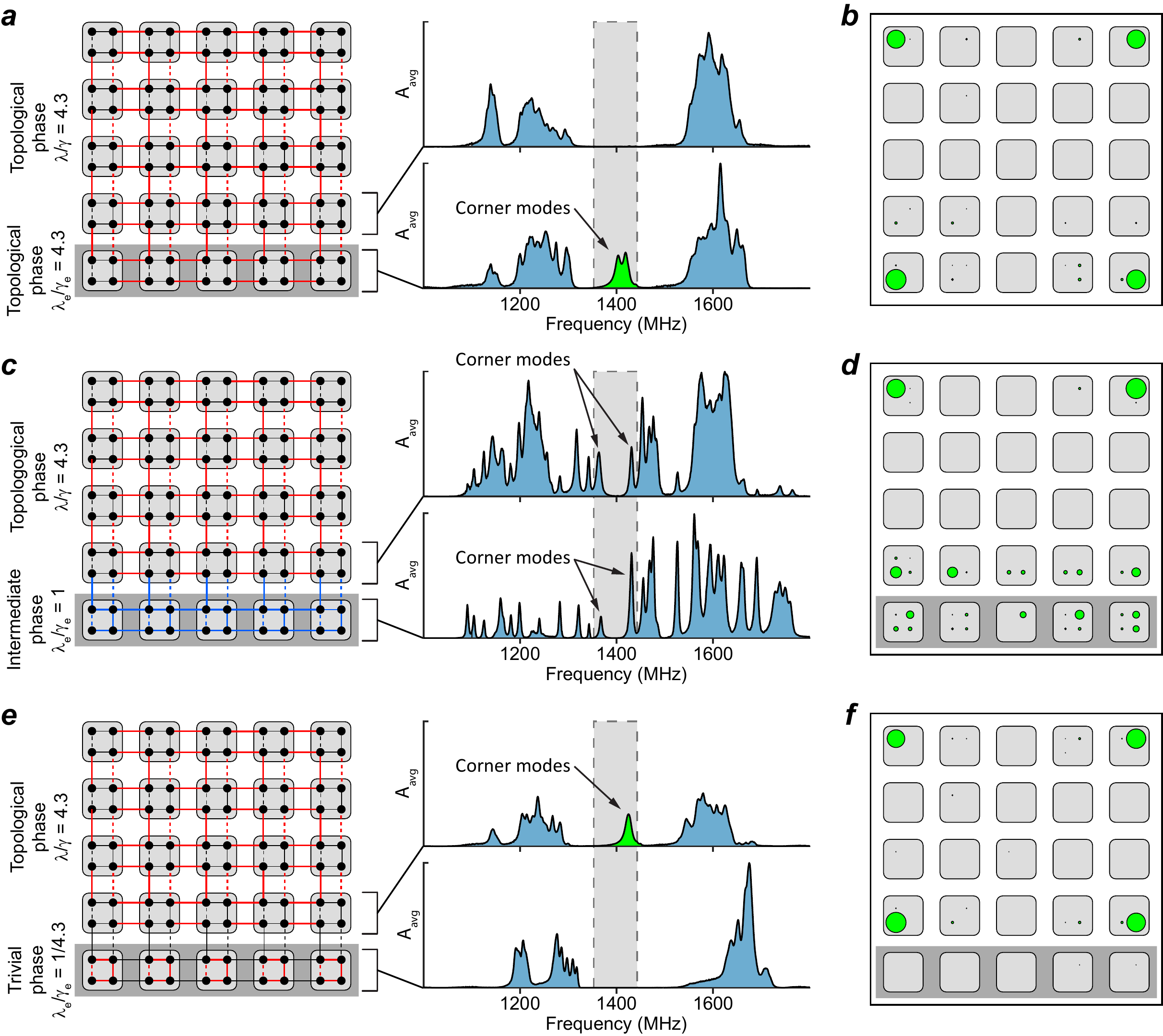}}%
		    \caption{
			{\bf Experimental test of topologically protected corner states during edge deformation --  }
			%
		    %
		    \textbf{(a)} The entire array is initially set in the topological phase with $\lambda / \gamma = 4.3$. The bottom two rows of unit cells display a bandgap, with mid-gap modes -- the topological corner modes -- appearing only on the bottom row.
		    \textbf{(b)} Measured spatial distribution of modes within the bandgap, summed over the shaded band in (a).
		    \textbf{(c)} The unit cells on the bottom edge are now set at a transition point between the topological and trivial regimes, with $\lambda_e / \gamma_e = 1$ (blue lines). The bandgap along the bottom edge narrows but remains open. Due to the finite size of the array, the corner modes couple to each other and their degeneracy is lifted.
            \textbf{(d)} Measured spatial distribution of modes within the bandgap, summed over the shaded band in (c). The in-gap modes are delocalized between the unit cells in the bottom two rows.
            \textbf{(e)} The unit cells in the bottom row are finally brought into the trivial regime with $\lambda_e / \gamma_e = 1/4.3$, while the rest of the array remains topological. The mid-gap modes are shifted one row up towards the new quadrupole topological phase boundary.
            \textbf{(f)} Measured spatial distribution of modes corresponding to (e). The mid-gap modes localize on the new corners of the quadrupole topological phase.
            }
            
	    \label{fig4}
	\end{adjustwidth}
\end{figure}%

To demonstrate that the corner-localized modes are not the result of local effects particular to the physical edges of the array, we tune the unit cells on the lowest row from the topological regime ($\gamma<\lambda$) to the trivial regime ($\gamma > \lambda$).
This experiment begins with the entire array in the original topological phase ($\lambda / \gamma \approx 4.3$) as shown previously in Fig.~\ref{fig3}. 
For this configuration, we plot the average absorptance spectra of the bottom two rows of unit cells separately, revealing that both rows are gapped but that the bottom row supports the mid-gap modes (Fig.~\ref{fig4}a). 
As previously shown, these mid-gap modes are localized on the corners of the array (Fig.~\ref{fig4}b).
Next, we adjust the coupling rates on the bottom row of unit cells to be equal, i.e. $\lambda_e / \gamma_e = 1$. 
This is achieved simply by replacing the coupling capacitors within the network.
This modification narrows the band gap of the bottom two rows (Fig.~\ref{fig4}c), and the two lower corner modes delocalize from the original corners into the surrounding unit cells (Fig.~\ref{fig4}d).
Due to the finite size of the experimental array, the corner modes couple to each other at this point in the process and their degeneracy is lifted.
Finally, we make the bottom edge unit cells trivial by setting $\lambda_e / \gamma_e \approx 1/4.3$, broadening the band gap to its original width (Fig.~\ref{fig4}e). 
Though the physical bottom edge of the array is now in the trivial regime, the corner modes are not destroyed but simply recede to the new topological phase boundary.
This experimental observation confirms that the corner modes are not just a surface artifact, but rather they are a unique manifestation of the bulk quadrupole topological phase.
In contrast, if the corner modes were generated from localized defects on the corners, or even if they arose as the end-states of edge-localized 1D topological dipole insulators, the mid-gap modes would disappear during the edge deformation.
%

%


\vspace{12pt}

This work showcases the first experimental evidence of a new and uncharted family of topological phases of matter. 
Our metamaterial implementation of the quadrupole topological insulator confirms the existence of the theoretically predicted corner modes \cite{benalcazar2017quad} and firmly establishes their origin from the bulk quadrupole topology. 
With an eye toward the future, we note that our reconfigurable microwave platform can readily support spatiotemporal modulation of both on-site energy \cite{EstepIEEE} and coupling rates \cite{Peterson}, prompting future experiments on dynamical topological phenomena including pumping processes and quenches. The stage is set for rapid advances at both the fundamental and device levels.  

\vspace{24pt}
\noindent
\textit{\textbf{Note added:}} During preparation of this manuscript we learned of two parallel efforts to realize quadrupole topological insulators. Imhof et al (Ref. \cite{Imhof_quad17}) demonstrate an electronic system at MHz frequencies. Serra-Garcia et al (Ref. \cite{Huber_quad17}) demonstrate a phononic system at kHz frequencies.
\vspace{24pt}

\vspace{12pt}

\section*{Methods}

\textit{Design of the quadrupole topological insulator lattice -- }
Each unit cell is fabricated individually on Rogers RT/duroid 5880 substrate, with $35~\mu$m thick copper on each side.
Within each unit cell, we select the coupling parameter $\gamma$ by connecting the resonators through a $0.1$~pF capacitor.
Between unit cells, we select the coupling parameter $\lambda$ by connecting the resonators through a $1$~pF capacitor.
The coupling rate between resonators is a sub-linear function of capacitance such that $\lambda/\gamma \approx 4.3$ within the system.
This ratio is determined empirically from the frequency separation between modes in the unit cell and inter-unit cell plaquette, which are proportional to $\gamma$ and $\lambda$ respectively.
The coupling capacitors also capacitively load the resonators, increasing their effective length and therefore reducing the resonance frequency.
For bulk resonators the capacitive loading is similar and does not affect the bulk spectral characteristics.
However, the reduced capacitive loading for edge and corner resonators is compensated (to match the bulk loading) by adding capacitance to ground of $0.6$ pF and $1.2$ pF to the edge and corner resonators respectively.
Further effects of capacitive loading are discussed in Supplement \S S2.
\vspace{12pt}

\noindent
\textit{Spectrum and eigenmode measurements --}
We measure the power absorptance spectrum at each resonator within the tested networks by means of 1-port reflection ($S_{11}$) measurements using a microwave network analyzer (Keysight E5063A). 
The reflection probe is composed of a 50 $\Omega$ coaxial cable terminated in a $0.1$ pF capacitor, which is contacted to each resonator at an anti-node.
The absorptance (ratio of absorbed power to incident power) of each resonator is calculated as $A = 1 - \big| S_{11} \big| ^2$.
We also define the average absorptance for an array of $N$ resonators as $A_{\textnormal{avg}} = \frac{1}{N} \sum_n^N A_n$, where $A_n$ is the absorptance of the $n^{\textnormal{th}}$ resonator.
In this calculation, we apply a minimum threshold to remove probe induced background absorption.
The eigenmodes of the unit cell and $2\times2$ array (Figures \ref{fig2} and \ref{fig3}) are also measured with a microwave network analyzer by means of 2-port transmission measurements ($S_{21}$). 
The measurement is performed using a pair of probes as specified above, with one probe used for stimulus and the other measuring response.
The $S_{21}$ transfer function at the resonant frequency thus produces a direct measurement for the amplitude and phase response for the corresponding eigenmode.

\vspace{12pt}
\section*{Acknowledgements}

The authors would like to thank Prof. Jennifer T. Bernhard for access to the resources at the UIUC Electromagnetics Laboratory. 
This project was supported by the US National Science Foundation (NSF) Emerging Frontiers in Research and Innovation (EFRI) grant EFMA-1627184.
C.W.P. additionally acknowledges support from the NSF Graduate Research Fellowship.
G.B. additionally acknowledges support from the US Office of Naval Research (ONR) Director for Research Early Career Grant.
W.A.B. and T.L.H. additionally thank the U.S. National Science Foundation under grant DMR-1351895.

\vspace{12pt}
\section*{Author contributions}

C.W.P. designed the microwave quadrupole topological insulator, performed the microwave simulations and experimental measurements, and produced the experimental figures. W.A.B. guided the topological insulator design and performed the theoretical calculations. T.L.H. and G.B. supervised all aspects of the project. All authors jointly wrote the paper.

\small{\bibliography{references.bib}}
\newpage

\newcommand{\beginsupplement}{%
        \setcounter{table}{0}
        \renewcommand{\thetable}{S\arabic{table}}%
        \setcounter{figure}{0}
        \renewcommand{\thefigure}{S\arabic{figure}}%
        \setcounter{equation}{0}
        \renewcommand{\theequation}{S\arabic{equation}}
        \setcounter{section}{0}
        \renewcommand{\thesection}{S\arabic{section}}%
}

\beginsupplement

\begin{center}
\Large{\textbf{Supplementary Information: \\ Demonstration of a quantized microwave quadrupole insulator with topologically protected corner states}} \\
\vspace{12pt}
\vspace{12pt}
\large{{Christopher W. Peterson}$^1$,
{Wladimir A. Benalcazar}$^2$,
\mbox{Taylor L. Hughes}$^2$,
{Gaurav Bahl}$^{3,\ast}$} \\
\vspace{12pt}
{\footnotesize{University of Illinois at Urbana-Champaign\\}}
{\footnotesize{$^1$Department of Electrical and Computer Engineering\\}}
{\footnotesize{$^2$Department of Physics\\}}
{\footnotesize{$^3$Department of Mechanical Science and Engineering\\}}
{\footnotesize{$^*$To whom correspondence should be addressed; bahl@illinois.edu}} \\
\vspace{12pt}
\end{center}


\normalsize

\section{Comparison of unit cell threaded with $\pi$ flux and 0 flux}
A unit cell of our quadrupole topological insulator is a square of four resonators threaded with $\pi$ flux, as illustrated in the main manuscript Fig.~\ref{fig2}a.
In Fig.~\ref{fig2}b, we show the measured eigenmodes of a single unit cell, which match well with the theoretically predicted modes. 
In this section we discuss all four eigenmodes of this system, and also establish a contrast against unit cells threaded with $0$ flux.
We find that, without flux, the modes differ significantly in their spatial distribution but that their energy spectra can be similar if $C_4$ symmetry is broken.
\begin{figure}[h!]
    \begin{adjustwidth}{-1in}{-1in}
		    \centering
		    \makebox[\textwidth][c]{\includegraphics[scale = 0.7]{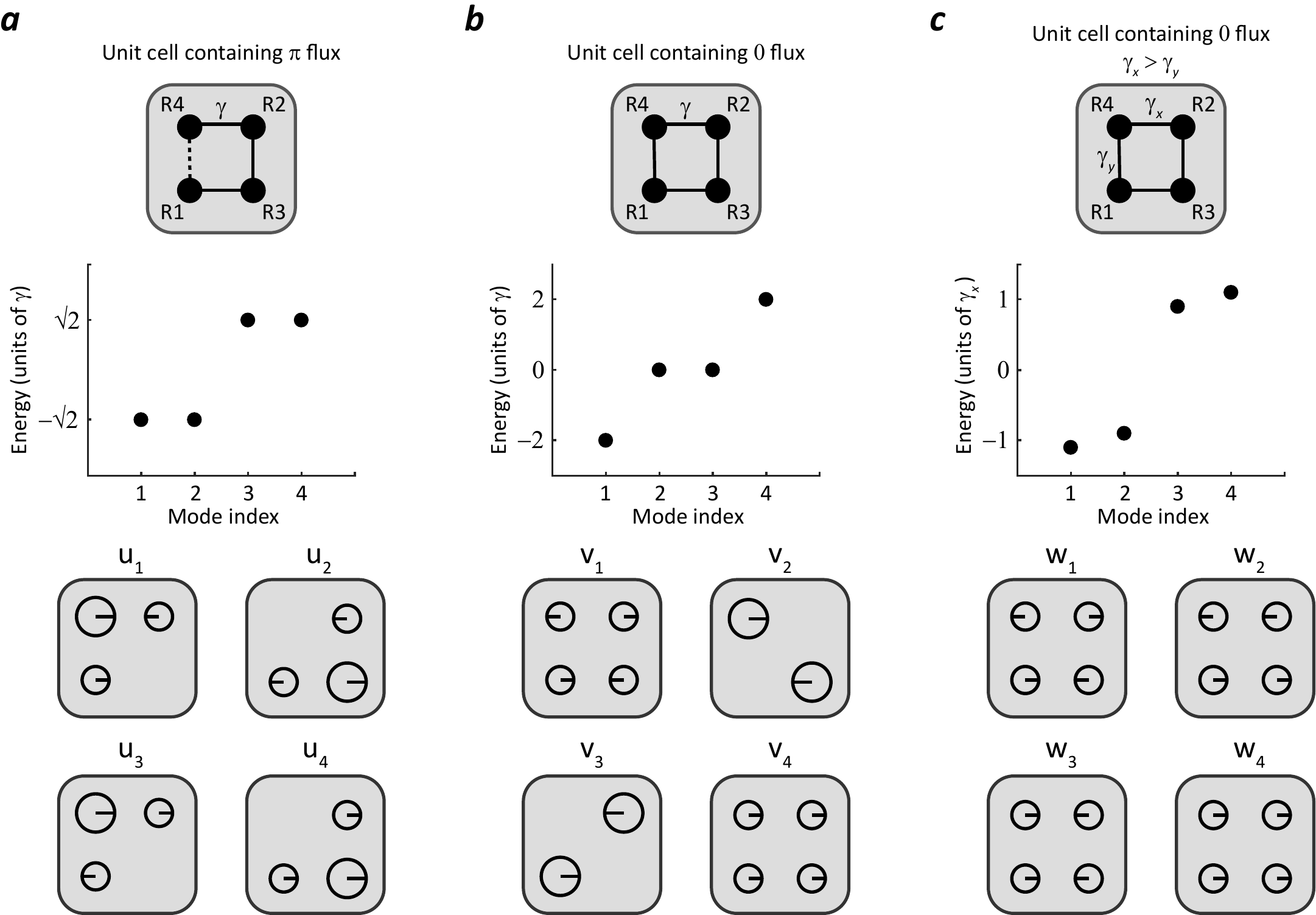}}%
		    \caption{
		    \textbf{Comparison of unit cells threaded with $\pi$ and 0 flux -- }
		    \textbf{(a)} Energy spectrum and eigenmodes of a unit cell with $\pi$ flux.
		    \textbf{(b)} Energy spectrum and eigenmodes of a unit cell with $0$ flux, with $\gamma_x = \gamma_y$.
		    \textbf{(c)} Energy spectrum and eigenmodes of a unit cell with $0$ flux, but having unequal coupling rates $\gamma_x > \gamma_y$. Energy separation between the lower pair of modes (and upper pair of modes) is proportional to $\gamma_y$.
            }
	    \label{figS1}
	\end{adjustwidth}
\end{figure}%
The calculated energy spectrum and eigenmodes of a unit cell with $\pi$ flux and equal coupling magnitudes $\gamma$ are shown in Fig.~\ref{figS1}a.
As described in the main manuscript, there are two pairs of degenerate eigenmodes.
These modes can be described by the orthonormal basis vectors 
\begin{equation}
\begin{split}
& u_1 = \begin{bmatrix} 1/2 & -1/2 & 0 & \sqrt{2}/2 \end{bmatrix} ~, \\
& u_2 = \begin{bmatrix} -1/2 & -1/2 & \sqrt{2}/2 & 0 \end{bmatrix} ~, \\
& u_3 = \begin{bmatrix} -1/2 & 1/2 & 0 & \sqrt{2}/2  \end{bmatrix} ~, \\
& u_4 = \begin{bmatrix} 1/2 & 1/2 & \sqrt{2}/2 & 0 \end{bmatrix} ~, 
\end{split}
\end{equation}
where each vector corresponds to the complex amplitudes of the resonators $\begin{bmatrix} \textnormal{R1} & \textnormal{R2} & \textnormal{R3} & \textnormal{R4} \end{bmatrix}$.
$u_1$ and $u_2$ are the degenerate pair of lower energy modes, and $u_3$ and $u_4$ are the degenerate pair of higher energy modes.
In the main manuscript Fig.~\ref{fig2}b we specifically measure the modes $u_1$, $u_2$.
Due to destructive interference arising from the $\pi$ flux within the plaquette, when one resonator is excited (here R3 or R4) the resonator on the opposite corner is not excited.
This property leads to the uniquely identifiable modes of this unit cell.
We also find that the location of the negative coupling does affect the relative phase between the resonators, leading to the opposite relative phase between resonators with and without negative coupling.
We can now contrast the above case against the calculated energy spectrum and eigenmodes of an identical unit cell having $0$ flux, as shown in Fig.~\ref{figS1}b.
These modes can be described by the orthonormal basis vectors
\begin{equation}
\begin{split}
& v_1 = \begin{bmatrix} 1/2 & 1/2 & -1/2 & -1/2 \end{bmatrix} ~, \\
& v_2 = \begin{bmatrix} 0 & 0 & -\sqrt{2}/2 & \sqrt{2}/2 \end{bmatrix} ~, \\
& v_3 = \begin{bmatrix} \sqrt{2}/2 & -\sqrt{2}/2 & 0 & 0 \end{bmatrix} ~, \\
& v_4 = \begin{bmatrix} 1/2 & 1/2 & 1/2 & 1/2 1/2 \end{bmatrix} ~.
\end{split}
\end{equation}
In this unit cell, only the modes $v_2$ and $v_3$ are degenerate, while $v_1$ has lower energy and $v_4$ has higher energy.
Since there is 0 flux threading the unit cell, when one resonator is excited, the resonator on the opposite corner is always excited as well.
While a unit cell with $0$ flux and identical horizontal and horizontal coupling rates ($\gamma_x = \gamma_y$) is not gapped, a bandgap can be opened by setting $\gamma_x > \gamma_y$ (i.e. breaking $C_4$ symmetry).
The calculated energy spectrum and eigenmodes for this case are shown in Fig.~\ref{figS1}c.
The modes can be described by the orthonormal basis vectors 
\begin{equation}
\begin{split}
& w_1 = \begin{bmatrix} 1/2 & 1/2 & -1/2 & -1/2 \end{bmatrix} ~, \\
& w_2 = \begin{bmatrix} 1/2 & -1/2 & 1/2 & -1/2 \end{bmatrix} ~, \\
& w_3 = \begin{bmatrix} 1/2 & -1/2 & -1/2 & 1/2 \end{bmatrix} ~, \\
& w_4 = \begin{bmatrix} 1/2 & 1/2 & 1/2 & 1/2 \end{bmatrix} ~.
\end{split}
\end{equation}
None of these modes are degenerate, but the lower pair (and upper pair) can be brought arbitrarily close for a large ratio $\gamma_x / \gamma_y$.
However, the spatial distribution of these eigenmodes clearly differs from a unit cell with $\pi$ flux, since all four resonators are equally excited in each mode.

\vspace{12pt}

\section{Microwave circuit implementation}
Here we discuss specifics of the circuit implementation for our quadrupole topological insulator design.
A transmission line model is provided and the connections between resonators are detailed.
We also discuss two cases of capacitive loading that are representative of situations encountered by resonators in our quadrupole topological insulator array.

\begin{figure}[h!]
	\begin{adjustwidth}{-1in}{-1in}
		    \centering
		    \makebox[\textwidth][c]{\includegraphics[scale = 0.6]{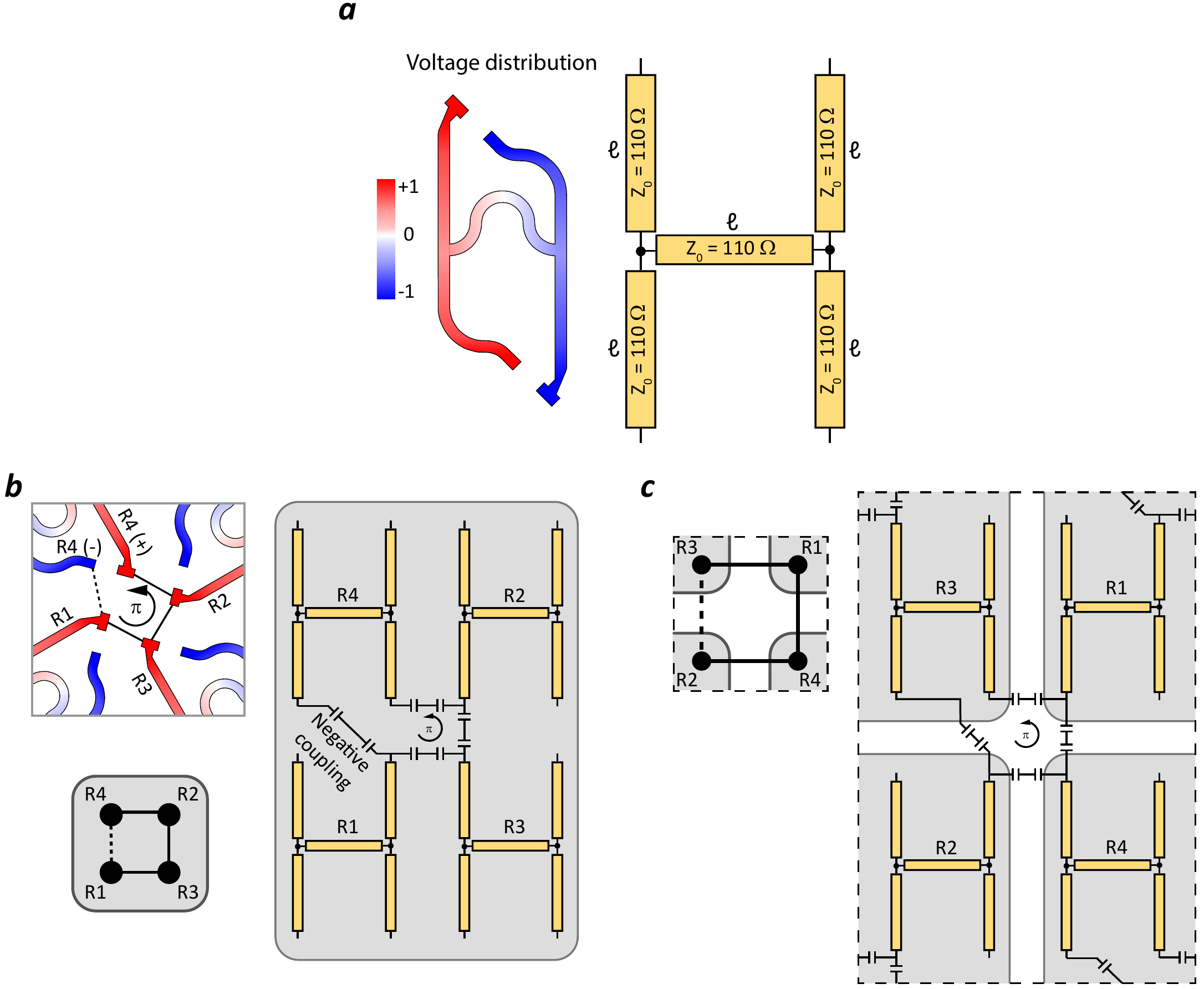}}%
		    \caption{
		    \textbf{Transmission line model -- }
		    \textbf{(a)} Transmission line diagram of an individual microstrip resonator. Each section is approximately the same length $\ell \approx 1.5$ cm and has the same characteristic impedance $Z_0 = 110~\Omega$, leading to a fundamental resonance frequency of $2.1$~GHz.
		    \textbf{(b)} The coupling that links resonators within the unit cell is implemented as two $0.2$ pF capacitors in series. The capacitors between R1 and R4 are connected to the out-of-phase anti-node of R4, creating $\pi$ flux threading the plaquette.
		    \textbf{(c)} The coupling that links resonators between unit cells is implemented as two $2$ pF capacitors in series. The capacitors between R2 and R3 are connected to the out-of-phase anti-node of R3 to produce the required $\pi$ flux.
            }
	    \label{figS2}
	\end{adjustwidth}
\end{figure}%

\subsection{Resonator design and coupling}
A transmission line representation of our resonator is shown in Fig.~\ref{figS2}a. 
The resonator is $H$-shaped, with the individual sections approximately the same length $\ell \approx$ 1.5 cm and width $w = 0.1$ cm.
The microstrip substrate is Rogers RT/duroid 5880 and the characteristic impedance of each section $Z_0 \approx 110~\Omega$.
This resonator design leads to an unloaded resonance frequency of approximately $2.1$~GHz (the measured resonance frequency is $2.08$~GHz).
To create a unit cell, four microstrip resonators are capacitively coupled as shown in Fig.~\ref{figS2}b.
Each capacitive coupling is implemented as two $0.2$ pF capacitors in series, resulting in a total coupling capacitance of $0.1$ pF between resonators within the unit cell.
The negative coupling is realized by connecting R1 to the opposite phase anti-node of R4.
The connections between unit cells are detailed in Fig.~\ref{figS2}c. 
Each capacitive coupling is implemented as two $2$ pF capacitors in series, for a total inter-cell coupling capacitance of $1$ pF.
The coupling rates $\gamma$ and $\lambda$ are extracted from the measured data in the limiting cases $\lambda \rightarrow 0$ and $\gamma \rightarrow 0$ respectively (main manuscript Fig.~\ref{fig2}).
We find that the ratio of the frequency separation between the degenerate mode pairs in these isolated intra-unit cell and inter-unit cell cases is approximately $4.3$, which implies the coupling rate ratio $\lambda / \gamma \approx 4.3$.

\subsection{Systematic and random disorder in the coupling rates}

\begin{figure}[h!]
    \begin{adjustwidth}{-1in}{-1in}
		    \centering
		    \makebox[\textwidth][c]{\includegraphics[scale = 0.45]{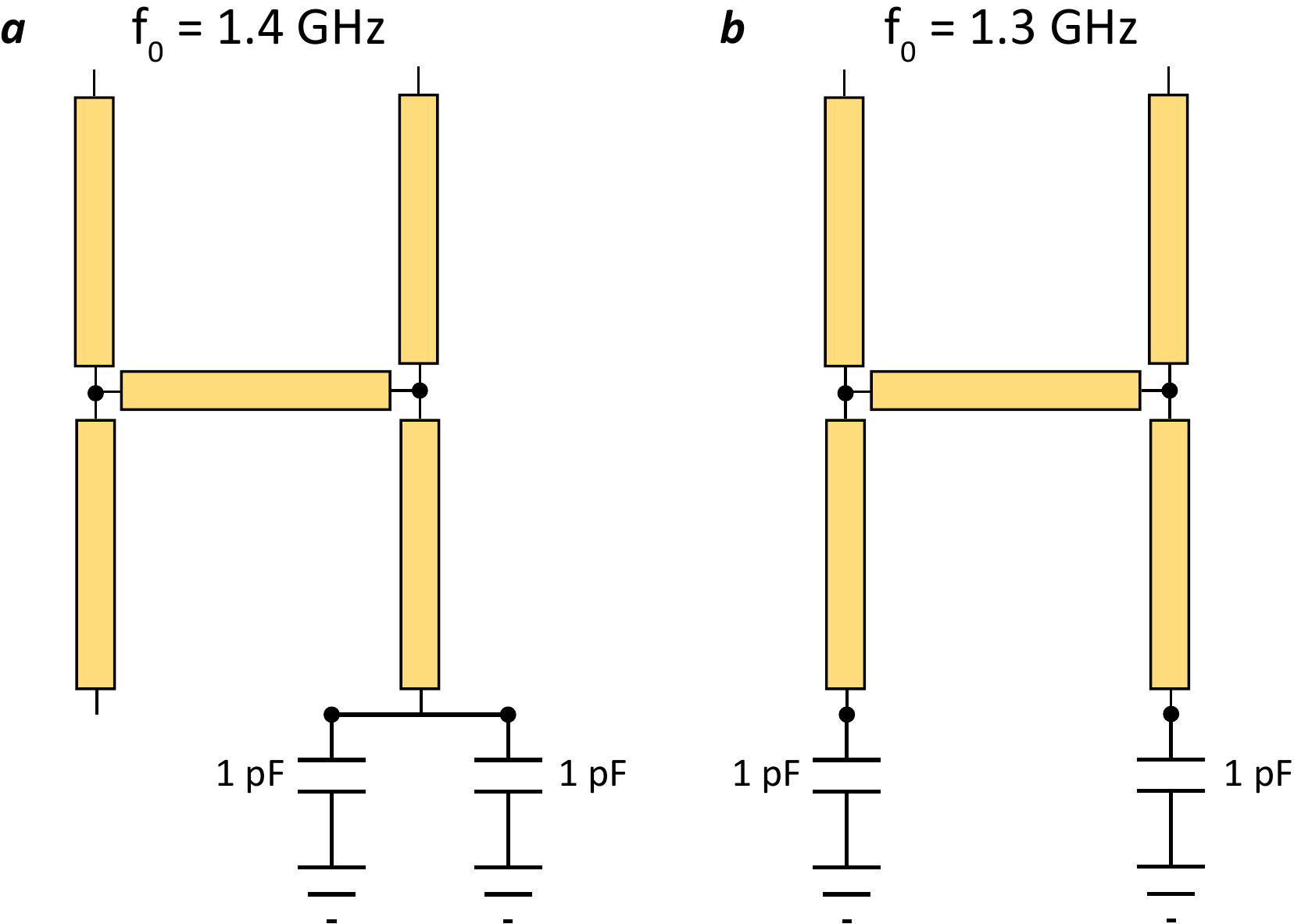}}%
		    \caption{
		    Comparison of resonators loaded with identical total capacitance. Resonance frequencies are calculated through simulation using Keysight ADS.
		    \textbf{(a)} A resonator with $2$ pF loading on a single arm: the resonance frequency is shifted from $2.1$~GHz to $1.4$~GHz due to the loading. This situation can be found for the intra-unit cell coupling of resonators R1, R2, R3 (Fig.~\ref{figS2}b) and for the inter-unit cell coupling of resonators R1, R2, R4 (Fig.~\ref{figS2}c).
		    \textbf{(b)} A resonator with $2$ pF loading distributed to two opposite polarity arms: the resonance frequency is shifted from $2.1$~GHz to $1.3$~GHz due to the loading. This situation can be found for the intra-unit cell coupling of resonator R4 and for the inter-unit cell coupling of resonator R3. 
            }
	    \label{figS3}
	\end{adjustwidth}
\end{figure}

Small differences in the capacitive loading of resonators inside our quadrupole topological insulator array implies disorder in both the resonance frequencies and coupling rates.
The impact of this disorder is seen in the measured eigenmodes in the limits $\gamma \rightarrow 0$ and $\lambda \rightarrow 0$ (main manuscript Fig.~\ref{fig2}).
In the full array, such disorder also results in the lower bulk band being split (Fig.~\ref{fig3}b).
We can understand how these differences in capacitive loading arise by examining two representative cases of resonators (Fig.~\ref{figS3}) loaded with identical total capacitance but with distinct spatial distributions.

In Fig.~\ref{figS3}a, we examine the case where both capacitors are on the same arm of the resonator. This is the case for the intra-unit cell coupling of resonators R1, R2, R3 (Fig.~\ref{figS2}b) and for the inter-unit cell coupling of resonators R1, R2, R4 (Fig.~\ref{figS2}c). The addition of $2$~pF capacitance to ground on one arm of the resonator causes a frequency shift to $1.4$~GHz (in contrast to the unloaded resonance at $2.1$~GHz).
In Fig.~\ref{figS3}b, we examine the case where the same $2$~pF capacitance is spread to two arms of the resonator. This is the case for the intra-unit cell coupling of resonator R4 and for the inter-unit cell coupling of resonator R3. 
Here, the resonance frequency shifts to $1.3$~GHz, i.e. lower than in the case where both resonators are on the same arm.
Although the total capacitance on each resonator is identical, these representative cases illustrate that the spatial distribution of capacitors impacts the degree of capacitive loading on the resonator.
Thus, systematic disorder in the capacitive loading of each resonator, which impacts both the coupling rate between resonators and their resonance frequencies, arises throughout our array.
In addition, the system also possesses randomized disorder due to manufacturing variation in the discrete component values.
Specifically, the $0.2$ pF capacitors have a tolerance of $\pm 0.05$~pF and the $2$~pF capacitors have a tolerance of $\pm 0.1$~pF. 

\end{document}